%% file: ICRC2023_I3Tilt.tex
\documentclass[a4paper,11pt]{article}

\usepackage{pos}
\usepackage{lipsum} 
\usepackage{wrapfig}

\title{An improved mapping of ice layer undulations for the IceCube Neutrino Observatory}

\ShortTitle{An improved mapping of ice layer undulations for the IceCube Neutrino Observatory}

\author{The IceCube Collaboration \\{\normalsize \normalfont(a complete list of authors can be found at the end of the proceedings)}\\}

\emailAdd{dmitry.chirkin@icecube.wisc.edu}
\emailAdd{martin.rongen@fau.de}

\abstract{

A precise understanding of the optical properties of the instrumented Antarctic ice sheet is crucial to the performance of the IceCube Neutrino Observatory, a cubic-kilometer Cherenkov array of 5,160 digital optical modules (DOMs) deployed in the deep ice below the geographic South Pole. We present an update to the description of the ice tilt, which describes the undulation of layers of constant optical properties as a function of depth and transverse position in the detector. To date, tilt modeling has been based solely on stratigraphy measurements performed by a laser dust logger during the deployment of the array. We now show that it can independently be deduced using calibration data from LEDs located in the DOMs. The new fully volumetric tilt model not only confirms the magnitude of the tilt along the direction orthogonal to the ice flow obtained from prior dust logging, but also includes a newly discovered tilt component along the flow.

\vspace{4mm}
{\bfseries Corresponding authors:}
Dmitry Chirkin$^{1}$, Martin Rongen$^{2*}$\\
{$^{1}$ \itshape 
Dept. of Physics and Wisconsin IceCube Particle Astrophysics Center, University of Wisconsin
}\\
{$^{2}$ \itshape 
Erlangen Centre for Astroparticle Physics, Friedrich-Alexander Universität Erlangen-Nürnberg
}\\
$^*$ Presenter

\ConferenceLogo{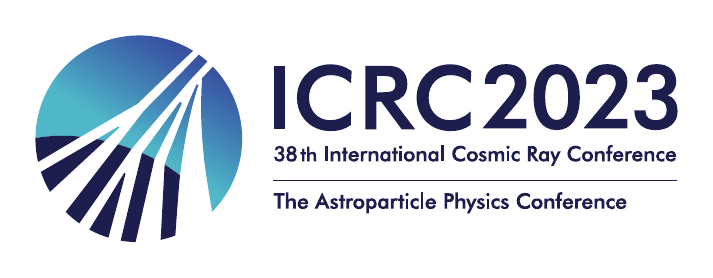}

\FullConference{The 38th International Cosmic Ray Conference (ICRC2023)\\ 26 July -- 3 August, 2023\\ Nagoya, Japan}
}

\begin{document}

\maketitle

\section{Introduction and previous ice layer undulation models}\label{sec1}

The IceCube collaboration has lowered a re-usable dust logger \cite{logger} in 8 of the drilled holes. The device shone a fan-shaped horizontal beam of laser light which was recorded by a downward-pointing Photo-Multiplier Tube (PMT) after scattering in the ice. Special baffles prevented the light from going directly from the laser to the PMT. The logger produced a precise record of dust layers in the ice with a resolution of $\sim 2$ mm in depth, sufficient to resolve narrow bands corresponding to prehistoric dust depositions. These features were matched between the 8 locations across the detector, and it was found that the depth of the features can vary by as much as 60 m within the 1 km$^3$ volume of IceCube. Henceforth we will refer to "ice layers" as layers with nearly identical optical scattering and absorption, as well as ice crystal density and fabric. In the context of IceCube ice models, the ice layers have so far been successfully described by averaging said properties in 10\,m depth increments at a lateral reference xy position. Thanks to the dust logger, it has long been clear that a model of ice layer "undulations" is needed because the ice sheet is not horizontally uniform. The layer undulations are understood to be a result of the bedrock topography only having been gradually smoothed out as the glacier accumulated, as commonly mapped out on larger distance scales using ground penetrating radar (i.e. \cite{CresisGPR}).

Initially, an ice layer "tilt model" was designed that described these undulations as a change in ice layer depth along a single "gradient" direction (1D tilt model), running almost precisely from NE to SW along the 45 degree grid North \cite{spiceMie}. The 1D tilt model was deemed accurate enough to be used for all analyses since 2013. The model linearly interpolates ice layer depths in between the dust logger locations (as projected onto the "gradient" direction) and linearly extrapolates them out into the space around the detector. Since 2013 we have discovered that the South Pole ice exhibits an optical anisotropy that affects photon propagation depending on their direction with respect to the "axis of anisotropy" that coincides to within $\sim 1^\circ$ with the direction of the ice flow\footnote{Surface ice at the South Pole moves from SE to NW along 135 grid North direction at a rate of around 10 m/year.} at the South Pole. It appears that more light propagates along the anisotropy axis than any other direction. This effect was first described in \cite{spiceLea}, and our understanding of it has recently significantly improved \cite{spiceBFR}. This more precise description of photon propagation has revealed that our simplified "gradient" tilt model should become our next target for improvement.

In this report, we describe the development of a 2D tilt model. 
Nearly all of the 5,160 IceCube optical sensors have 12 working LEDs that were flashed individually (100-200 flashes per LED), resulting in a set of more than 60,000 unique flasher patterns, each with hundreds of nearby DOMs with recorded light. This rich calibration set provides enough data to calculate significantly improved tilt corrections. 
The data reveal additional components to the tilt model beyond the 1D gradient model.

\section{New 2D model of ice layer undulations}\label{sec2}

\begin{wrapfigure}{r}{0.4\textwidth}
\vspace{-20pt}
\includegraphics[width=\linewidth]{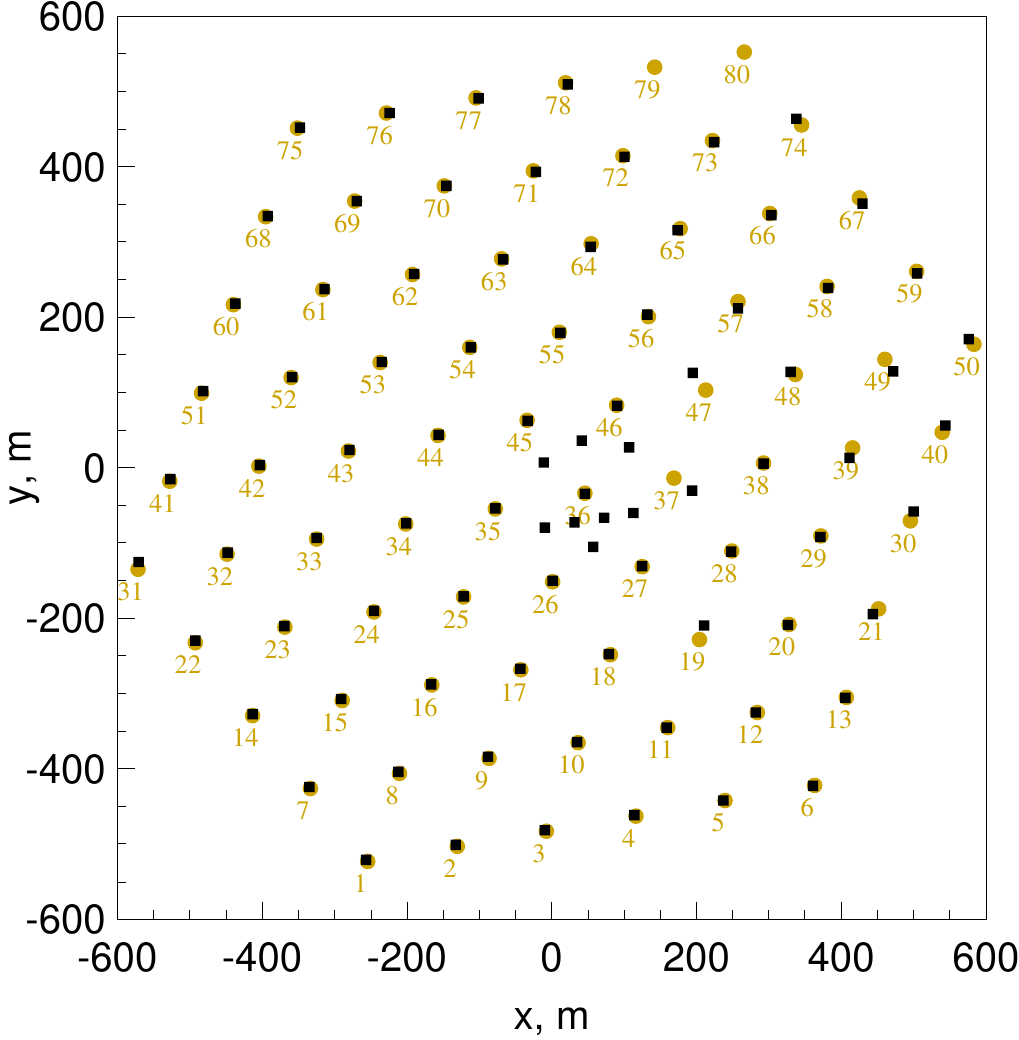}
\caption{Lateral geometry of the IceCube detector. Black squares denote the string locations. Golden circles mark the grid locations of the new tilt parameterization. }
\label{geo}
\vspace{-20pt}
\end{wrapfigure}

We have built a fully-2D parameterization of the ice tilt as follows. We first define 80 locations on a regular hexagonal grid in x and y to match 78 of the string locations of the IceCube detector as best as possible (see Fig.\ \ref{geo}). At each of the 80 xy locations we define a depth table of 125 ice layer tilt corrections spaced out by 10 m in depth, covering the full depth range of the deployed IceCube sensors and extending above it since we have dust logger measurements above the detector. We define the reference point of the ice table of optical properties to be at the grid point closest to string 36 near the center of the IceCube detector. Here, by definition, the ice layers need no tilt correction and the values in the tilt table are set to 0. This leaves at most (80-1)*125=9875 remaining tilt corrections as free parameters in the tilt table.

To compute the tilt corrections at a given xy location within the detector hexagon, we linearly interpolate between the closest 3 grid points (which form an equilateral triangle) for each ice layer in the depth table. Then, for a given depth, a linear interpolation between the nearby depth points is performed. To extrapolate to xy locations outside of the detector, the hexagon outline of the xy grid is scaled radially out until it intersects with the given xy point, matching the tilt correction to the value assumed on the hexagon boundary. This results in tilt values outside the detector that are well-behaved (bounded). Such construction results in tilt correction being a continuous function of x, y, and z, albeit not necessarily smooth at the boundaries between local grid triangles. However, since the tilt varies slowly across the detector volume, this non-smoothness has been deemed an acceptable compromise in exchange for the simplicity and numerical speed of the tilt estimation.

Before embarking on a fit simultaneously deducing the tilt correction at each grid location as well as the ice properties at the reference location, the sensitivity of the LED flasher data to small tilt corrections was established in a simplified initial fit. For this the tilt correction local to each individual DOM was deduced by globaly shifting the existing ice table of optical properties by small amounts up and down until the best description of the calibration data of this DOM acting as emitter was achieved. 
Since no correlation between neighboring DOMs is assumed, these initial tilt corrections can fluctuate within their resolution of 1-2\,m. The generate a smoother tilt map as starting point for the full fit, the initial tilt corrections were 3D smoothed within a 150m radius followed by a further Savitzky-Golay filter for all DOMs along each string. The tilt correction values at tilt table grid points were then obtained by linear interpolation from the closest 3 strings, usually influenced mainly by just the closest string. At this point the best-fit tilt corrections at the reference xy grid point were nonzero. To ensure the tilt values are 0 at the reference xy grid point we needed to re-sum the optical properties of the shifted table into the nominal depth table. 

For the full fit, the ice parameters (scattering, absorption, and ice crystal density) in 171 ten-meter layers at grid point 36 were added to the pool of free parameters, raising the total to $125\cdot (80-1)+3\cdot 171=10388$. The fitting procedure was broken down into multiple iterations, where sets of 100-1,000 random variations of either ice-only or both ice and tilt parameters were simulated and compared to the LED calibration data. The ice parameters were sampled from a uniform distribution of width $\pm 10$\%, then $\pm 5$\%, and finally $\pm 3$\%. The tilt variations were also drawn from a uniform distribution of width $\pm 4$ m, then $\pm 2$ m. These were then smoothed in two steps since we expect ice layers to vary smoothly over a large area, rather than varying up and down between nearby grid positions. First we minimized the square of the deviation of a grid point value from the average of the closest (up to 8) neighbors, summed over all 10,000 grid points (including the nominal points at the reference grid location). Second, we minimized the sum of the squares of the differences between consecutive (in depth) tilt corrections along each lateral xy grid location to make the layers vary smoothly in thickness, staying close to the nominal 10 m.

\begin{wrapfigure}{r}{0.43\textwidth}
\includegraphics[width=\linewidth]{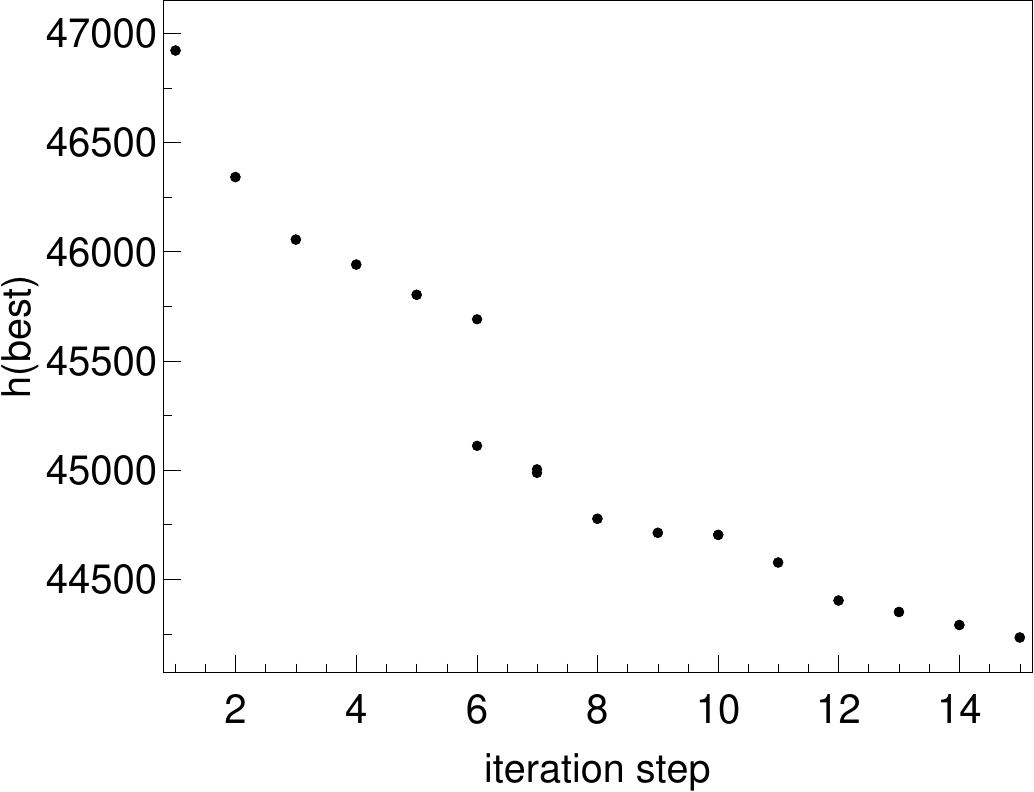}
\caption{Progression of the best fit values of the comparison function versus fit iterations.}
\label{fit_progress}
\end{wrapfigure}
Each parameter variation was simulated with all 60,000 LED configurations available in our calibration data set. For each configuration 100-250 individual flashes are available in data. Due to computing time constraints, for each configuration only 10 flashes were simulated. The data and simulation were compared using the approach of \cite{llhpaper}, which takes into account Poisson fluctuations in both data and simulation, as well as possible mis-modeling of data with simulation, a difference that would persist even at the limit of infinite statistics. Mis-modeling may occur due to inaccuracies in our description of the calibration LED events, or due to simulation simplification/acceleration approximations, such as sensor oversizing; we estimate this effect to be 10\%. The value of the comparison function (which can be thought of as "distance" between data and simulation) is denoted with $h_i$, where $i$ is an ice model realization.

We assume that we can describe resulting $h_i$ from ice variations with a paraboloid (second-order Taylor expansion) around the best solution. This paraboloid was fitted to the simulated ice model variations at each iteration for a range of regularization parameters discussed in the next paragraph, resulting in a set of "proposed" ice solutions. These were then re-simulated and added to the ensemble of ice variations, allowing us to re-fit the paraboloid several times and improve the best solution found at each iteration. The unknowns of the fit are the ice-only or ice+tilt parameters $b_n$, and the components of the (up to) 10513 by 10513 {\it curvature} matrix (second-order coefficients).

This is a highly under-constrained problem since we are trying to fit $\gtrsim 10^8$ unknown parameters to $\sim 10^2-10^3$ ice model variations. Thus, we had to impose some regularizations, further simplifications, and other conditions. A standard approach to such an under-constrained problem is to find a solution with the minimum norm of the solution. The solution can be chosen to make the paraboloid go directly through the set of simulated models, or with additional regularizations, reducing the problem to a matrix inversion. We tried a number of such "direct inversion" approaches but found the fitted paraboloid is virtually never positive-definite, with the ice solution lying in a saddle point rather than at the minimum. We thus explicitly enforce the positive-definite nature of the curvature matrix. To do this, we represent the curvature matrix as a product of a matrix and its transpose, $c_{nm}=a_{kn}a_{km}$\footnote{The convention that summation over the repeated index is implied is used throughout this report.}. Additionally, we add regularizations to require that the norm of the curvature matrix be low (so the paraboloid is as flat as possible to avoid over-fitting), and terms that describe the tilt map smoothing conditions. Similarly, we impose second derivative smoothness conditions on scattering, absorption, and ice crystal density. Finally, in the initial few iterations we also impose a regularization on the norm of the ice solution vector so that it is constrained to the area in the parameter space sampled with the ice variations. Altogether, these conditions can be written down as the following function, to be minimized:
\begin{multline}
L=\sum_{i=1}^{N_\mathrm{sim}}(Q^i+C_0-h_i)^2+\alpha\cdot(\sum_{k,n=1}^{N_\mathrm{par}}a_{kn}^2)^2+\beta\cdot(\sum_{k=1}^{N_\mathrm{par}}(D_{kn}(b_n+T_n))^2)^2+\gamma\cdot(\sum_{n=1}^{N_\mathrm{par}}b_n^4) \quad,\\
\text{where}\quad Q^i=\sum_{k=1}^{N_\mathrm{par}}(a_{kn}(\Delta_n^i-b_n))^2\quad.
\label{eqL}
\end{multline}
Here $i$ indexes the ice variations (out of $\sim 10^2-10^3$ realizations); $h_i$ is the comparison function (distance) between data and simulation for ice variation $i$; $k,n$ are the indices in the ice parameter space (1...10513); $\Delta_n^i$ are the components of the ice variation $i$, measured either from the best fit of the previous iteration, or from the initial solution; and $b_n$ is the best/fitted ice of this iteration, for which $L$ reaches its minimum. $Q^i+C_0$ describes the fitted paraboloid, with $C_0$ being the constant giving the best value of the paraboloid, and $Q^i$ being the quadratic form.

The curvature regularization strength $\alpha$ is optimized for each iteration. The three correlation terms describing ice and tilt smoothness constraints are described with the term starting with regularization strength $\beta$ (index running over 1,2,3 is omitted for brevity). $D_{kn}$ describes the specific regularization construction (such as the difference between consecutive tilt values at each xy grid location), and $T_n$ describes the actual starting ice parameter values, which are necessary to calculate the ice parameters $b_n+T_n$ from the solution vector $b_n$ (which described the variation from the starting ice model).

In order to make the values of the ice parameter solution vector $b_n$ commensurate between the tilt and ice components, the tilt components were measured in hectometers (100 m), and ice components were taken as natural log of their table values. This reduces the expected scale of the components of the solution vector $b_n$ to the order of $\sim$ 0.01. We have eventually introduced additional small correction to the tilt scale (factors $\sim 2$), as it became clear that the matching between the scale of the ice and tilt components could be improved further (as gauged by the speed of convergence to the best solution).

Minimization of function $L$ in Eq.\ \ref{eqL} was performed with a Newton-Rapson gradient descent method. As a starting point we set $b_n=0$, $a_{kn}=\xi\delta_{kn}$, reducing $L$ to a function of just one variable, $\xi$, which is easily solved for $\xi$. The gradients of $L$ with respect to $b_n$ and $a_{kn}$ can be easily calculated, and a search for the best solution along the gradient reduces to an 8th order polynomial equation. We found that an efficient search can be performed by splitting the gradient descent into two steps: first search along the gradient with respect to $a_{kn}$, followed by $b_n$. The gradient search with respect to $C_0$ can be solved analytically. These steps are repeated $\sim 1000-5000$ times until the desired convergence is achieved.
At every search along the gradient $\partial L/\partial b_n$, components of this gradient, which correspond to the tilt components at the reference location at grid point 36, are set to 0, thus fixing these to their nominal zero values.

We next focus our attention on the calculation of the covariance matrix, which is related to the matrix $a_{kn}$ (it is the inverse of the curvature matrix $c_{nm}$). We found that the off-diagonal elements of the correlation matrix were not well constrained, and observed a "noise floor" in the off-diagonal elements that extended throughout the entire matrix, leading to non-physical long-distance correlations such as between elements describing ice and tilt components at shallow vs. deep locations in the detector. Thus, we decided to explicitly reduce the number of non-zero off-diagonal elements in the matrix $a_{kn}$ to a smaller set, which not only stabilized the matrix $a_{kn}$ and the resulting correlation matrix, but also accelerated the calculation by several orders of magnitude.

\begin{wrapfigure}{r}{0.45\textwidth}
\vspace{-5pt}
\includegraphics[width=\linewidth]{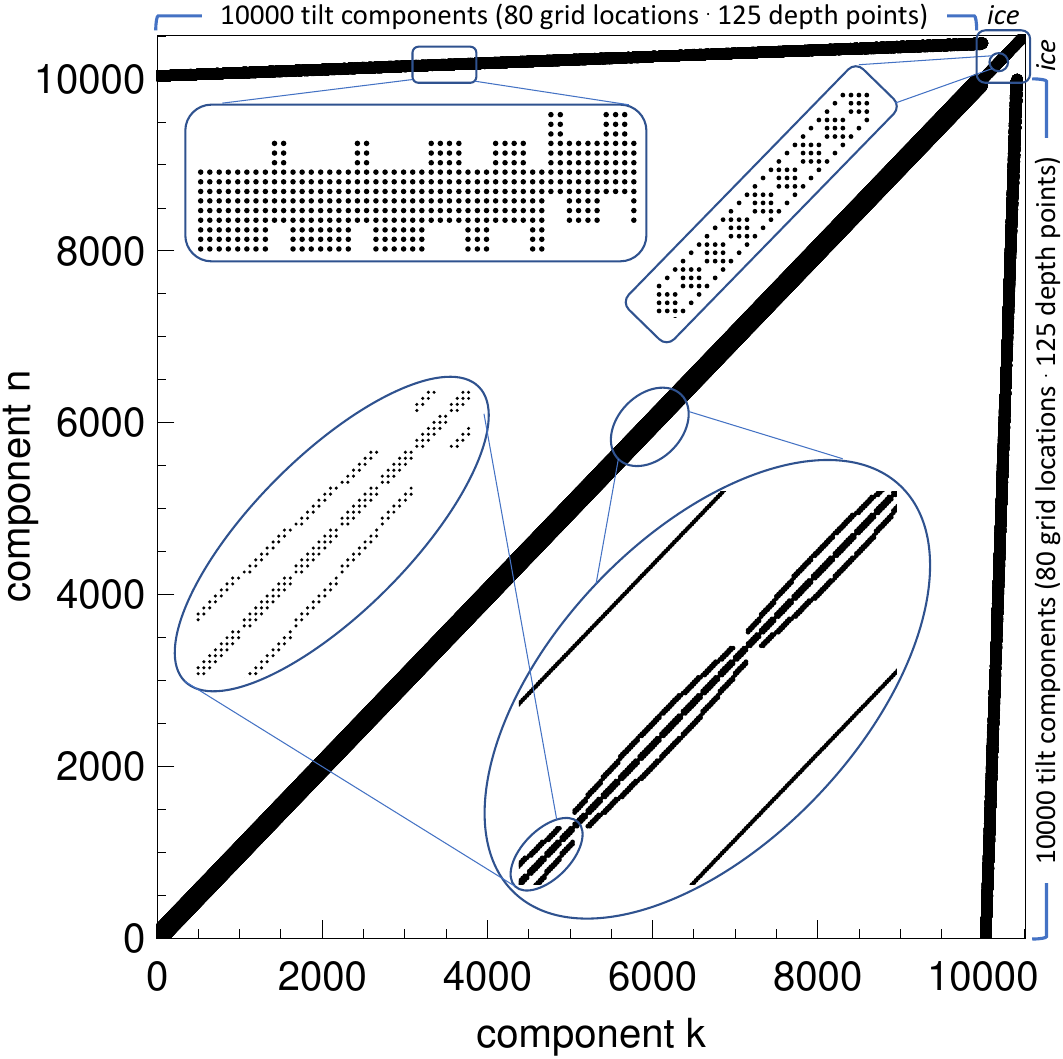}
\caption{Allowed elements of the matrix $a_{kn}$.}
\vspace{-10pt}
\label{sparse}
\end{wrapfigure}
The set of "allowed" elements of the matrix $a_{kn}$ was defined as follows (see Fig.\ \ref{sparse}). First, all diagonal elements were included. Next, ice components that relate any of the scattering, absorption, and ice crystal density depth elements to the depth layers above and below, and to each other, were also included. The tilt components between tilt elements immediately above or below, and immediately to the (up to 6) side neighbors were included. Finally, the tilt components were allowed to correlate to the ice components in the layer (or nearby layers) for which the tilt component defined the ice layer tilt. Because the curvature matrix is a product of $a_{kn}$ and its transpose, effectively all "allowed" correlations are doubled in length, e.g., ice properties between layers separated by one layer in between may result in non-zero correlation values. We did observe that this model might be limiting to the optical properties of the ice layers, and may add allowed (non-zero) elements in that part of the matrix in the future. However, the elements involving tilt components (tilt-tilt or tilt-ice) allowed by the description above, appear to be more than sufficient. This is likely because tilt components are highly constrained by smoothness regularizations, and thus correlations between neighboring elements can absorb correlations with elements that are farther away.

\section{Uncertainty estimation and ice model sampling}\label{sec3}

We have so far avoided calling the quantity $h_i$, comparing photon hits in data and simulation for a specific ice model $i$, a likelihood function, because, at the very least, it is not a function of data and the ice model. Every time it is evaluated, a new set of simulation is produced. When evaluated several times it forms a distribution with a mean ($\sim 44240$) and a Root Mean Squared (RMS), which we calculated for the best model to be $\sim 7.6$. To estimate uncertainties on the ice parameters we continue with this choice of likelihood-free inference. Here we describe our calculation based on the Approximate Bayesian Computation (ABC) method. The idea is to sample ice models that produce simulations which are sufficiently close to the data, gauging by the "distance" $h_i$. Ideally we should be sampling possible data realizations, however we note that $h_i$ is symmetric between data and simulation, with one difference: we simulate fewer events than we have in data, thus our calculation here will result in a conservative over-estimation of uncertainties.

Because the ice models that we will consider in the uncertainty calculation will result in $h_i$ that are not much higher than the values at the best model ($44210 \pm 7.6$), and since $h_i$ is a measure of a goodness-of-fit (calculated as a sum of $\sim 60,000$ LED configurations), we think it reasonable to approximate the size of $h_i$ fluctuations for all considered ice models with the same RMS value of 7.6. Furthermore, we approximate the distribution of $h_i$ with a Gaussian $\mu+G(\sigma)$ with $\sigma=7.6$ and a mean of $\mu=C_0+(A\cdot\delta x)^2$, $A$ being the matrix form of $a_{kn}$ and $\delta x$ being the vector form of the ice model deviation from the best fit $b_n$. The ABC method prescribes that we sample the space of ice parameters and accept those that satisfy the condition
\begin{equation}
h_i\sim G(\sigma)+C_0+(A\cdot\delta x)^2<F
\end{equation}
for some $F$, where $F$ should be small enough to only allow simulation instances that are sufficiently similar to data. Setting $\xi=A\cdot\delta x$, we start by sampling the space of ice model parameters by sampling $\xi$ from a scaled normal distribution $G(\alpha)$ for some $\alpha$, determined below. This results in the sum $(A\cdot\delta x)=\alpha^2\cdot\sum_N G(1)^2$ behaving as a $\chi^2$ distribution with a mean of $\alpha^2 N$ and RMS of $\alpha^2 \sqrt{2N}$. Since $N\lesssim 10513$ is very large, the value of this sum is close to just being $\alpha^2 N$ with a small ($\sim 1$\%) uncertainty. If we choose $F=C_0$, then we just need to sample the ice model space such that $(A\cdot\delta x)^2<|G(\sigma)|$. Here we switched to the absolute value of the Gaussian, since the left side is a sum of squares and as such cannot be negative. We now introduce the last approximation of this calculation, which also can only lead to an over-estimation of the ice parameter uncertainties: instead of the inequality we will sample the ice parameters to satisfy an equality $(A\cdot\delta x)^2=|G(\sigma)|$. This can be achieved by sampling $\alpha$ from the following function of the half-Gaussian distribution $|G(\sigma)|$:
\begin{equation}
\alpha=\sqrt{\sigma \cdot |G(1)|\over N} \quad .
\label{alfa}
\end{equation}

To sample the ice models from the parameter space so constructed, we first sample $\alpha$ using Eq.\ \ref{alfa}. Then we sample N variables $\xi$ from the Gaussian distributions $G(\alpha)=\alpha\cdot G(1)$. Finally we invert the equation $\xi=A\cdot\delta x$ to obtain the variation vector of parameters of the ice model $\delta x$. We can simplify this procedure by first decomposing $A^TA$ with a Cholesky transformation, obtaining the curvature matrix representation where $A$ is upper triangular. This allows to solve the system of equations $\xi=A\cdot\delta x$ by working backwards from the the highest components down ($x_N=\xi_N/a_{NN}$, then $x_{N-1}=(\xi_{N-1}-a_{N-1,N}\cdot \xi_N)/a_{N-1,N-1}$, etc.). Such an approach also simplifies fixing the tilt components at the reference grid location to 0. To achieve this we shuffle the elements of the curvature matrix to stack the components we want to keep fixed at the end, before performing the Cholesky decomposition. Then, when solving the system of equations $\xi=A\cdot\delta x$ we set the elements of $\xi$ with the highest indices to 0 (instead of sampling from scaled Gaussian distribution, which we continue to do for the rest of the components), which necessarily results in the corresponding elements of $\delta x$ being 0s as well. Then we re-shuffle the components of $\delta x$ back to to their original order and we get the correct sampling of the ice model parameters while keeping the tilt components at the grid point closest to string 36 fixed.

Finally, we note that in addition to calculating the covariance matrix from sampled ice models, we can calculate it directly by inverting matrix $A^TA$ (using the same Cholesky decomposition) and scaling the result with the average
\begin{equation}
<\alpha^2>=<|G(1)|>\cdot {\sigma\over N}=\sqrt{2\over\pi}\cdot {\sigma\over N} \quad .
\end{equation}
When inverting smaller subsets of the curvature matrix $A^TA$ (e.g., keeping all of the tilt components fixed) we found that applying a Singular Value Decomposition (SVD) to matrix $A$ itself and calculating the correlation matrix from there works just as well, and allows one to investigate the singular values of the correlation matrix to better gauge its convergence. This allowed us to discover that adding small amounts of regularization terms to the curvature matrix $A^TA$ in the form of $\beta^* \cdot D^TD$ ($D$ being the matrices from the regularization terms in Eq.\ \ref{eqL}) stabilized the inversion process to where the sampling procedure described here could be used reliably.
\begin{wrapfigure}{r}{0.4\textwidth}
\includegraphics[width=\linewidth]{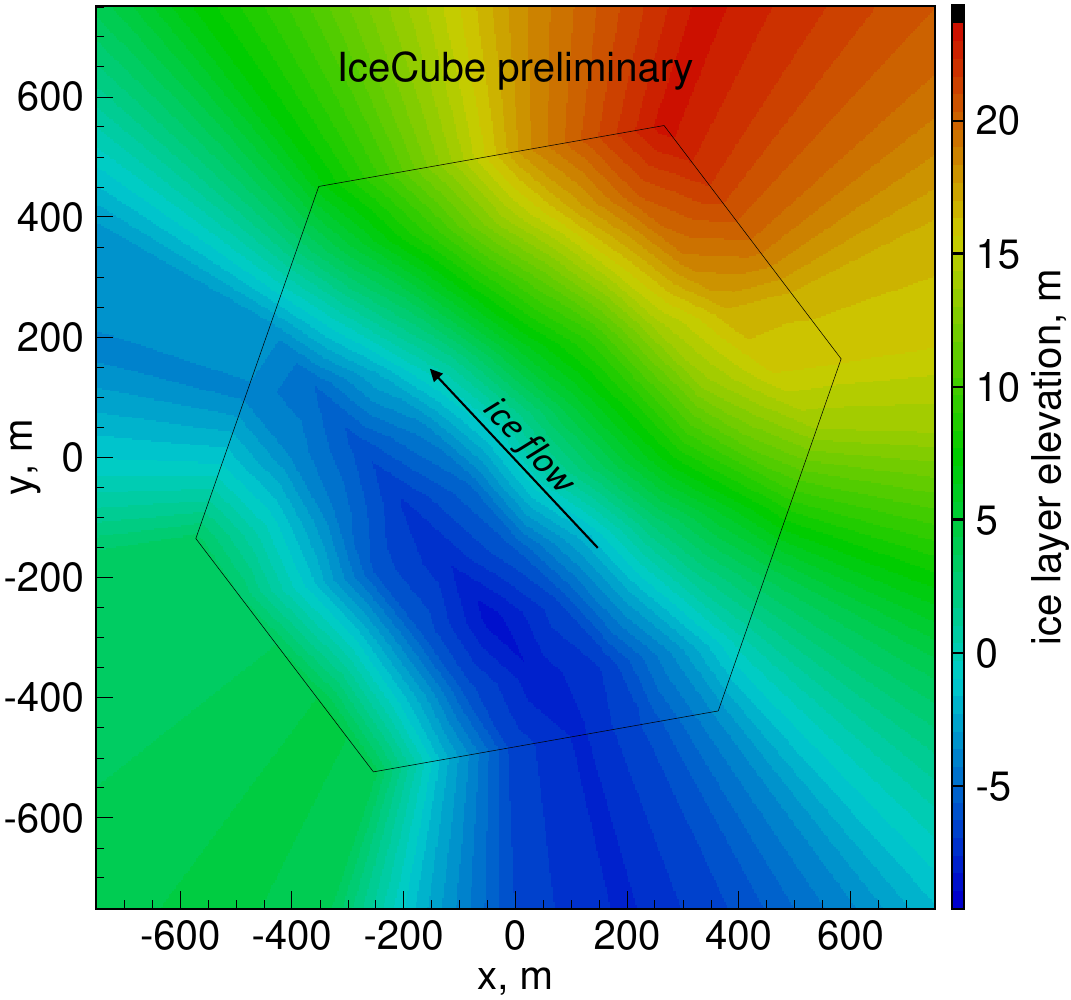}
\caption{Example slice through the deduced volumetric map of ice layer undulations. Shown is the elevation of an ice layer, defined at a depth of 2248\,m at the nominal position close to the center.}
\label{tiltmap}
\vspace{-30pt}
\end{wrapfigure}
Our first attempt at re-simulating the sampled ice models resulted in re-simulated values of $h_i$ that were much higher than predicted by the sampler. At that time we've had around 4000 simulated ice model samples that were used in the fit. Currently we are re-simulating proposed (at $5\sigma$ level) samples, and adding them into the fit. We find that the disagreement between proposed and simulated $h_i$ values is shrinking, and expect to reach an acceptable agreement after adding another $\sim$2000 samples.

\section{Result}\label{sec4}
The effort described in this report results in an improved description of ice layer undulations to $\sim 0.2$\,m, and ice layer properties (scattering, absorption, ice crystal density) to $\sim 0.5$\% statistical-only uncertainties. The full uncertainties including systematics remain to be re-evaluated. As evident from the example in Fig. \ref{tiltmap} the new layer undulations show significant deviations from the previous 1-D gradient assumption. Depth-averaged this in particular entails a 7\,m slope along the flow direction over the extend of the detector. Peculiarly the layer elevations are higher downstream, indicating that the flow locally follows an uphill bedrock topology. This requires further investigations including available ground-penetrating radar data. Our measure of the goodness of fit (see Fig.\ \ref{fit_progress}) for the full set of LED calibration data has also improved by an amount commensurate with the improvement due to the recent re-interpretation of the ice optical anisotropy~\cite{spiceBFR}.

\bibliographystyle{ICRC}
\bibliography{ICRC2023_I3Tilt}

\clearpage

\input{authorlist_IceCube.tex}

\end{document}

%% file: authorlist_IceCube.tex
\section*{Full Author List: IceCube Collaboration}

\scriptsize
\noindent
R. Abbasi$^{17}$,
M. Ackermann$^{63}$,
J. Adams$^{18}$,
S. K. Agarwalla$^{40,\: 64}$,
J. A. Aguilar$^{12}$,
M. Ahlers$^{22}$,
J.M. Alameddine$^{23}$,
N. M. Amin$^{44}$,
K. Andeen$^{42}$,
G. Anton$^{26}$
% ...
\\
\\
% ...
$^{12}$ Universit{\'e} Libre de Bruxelles, Science Faculty CP230, B-1050 Brussels, Belgium \\
% ...
$^{17}$ Department of Physics, Loyola University Chicago, Chicago, IL 60660, USA \\
$^{18}$ Dept. of Physics and Astronomy, University of Canterbury, Private Bag 4800, Christchurch, New Zealand \\
% ...
$^{22}$ Niels Bohr Institute, University of Copenhagen, DK-2100 Copenhagen, Denmark \\
$^{23}$ Dept. of Physics, TU Dortmund University, D-44221 Dortmund, Germany \\
% ...
$^{26}$ Erlangen Centre for Astroparticle Physics, Friedrich-Alexander-Universit{\"a}t Erlangen-N{\"u}rnberg, D-91058 Erlangen, Germany \\
% ...
$^{40}$ Dept. of Physics and Wisconsin IceCube Particle Astrophysics Center, University of Wisconsin{\textendash}Madison, Madison, WI 53706, USA \\
% ...
$^{42}$ Department of Physics, Marquette University, Milwaukee, WI, 53201, USA \\
% ...
$^{44}$ Bartol Research Institute and Dept. of Physics and Astronomy, University of Delaware, Newark, DE 19716, USA \\
% ...
$^{63}$ Deutsches Elektronen-Synchrotron DESY, Platanenallee 6, 15738 Zeuthen, Germany  \\
$^{64}$ Institute of Physics, Sachivalaya Marg, Sainik School Post, Bhubaneswar 751005, India \\
% ...

\subsection*{Acknowledgements}

\noindent
The authors gratefully acknowledge the support from the following agencies and institutions:
USA {\textendash} U.S. National Science Foundation-Office of Polar Programs,
U.S. National Science Foundation-Physics Division,
U.S. National Science Foundation-EPSCoR,
Wisconsin Alumni Research Foundation,
Center for High Throughput Computing (CHTC) at the University of Wisconsin{\textendash}Madison,
Open Science Grid (OSG),
Advanced Cyberinfrastructure Coordination Ecosystem: Services {\&} Support (ACCESS),
Frontera computing project at the Texas Advanced Computing Center,
U.S. Department of Energy-National Energy Research Scientific Computing Center,
Particle astrophysics research computing center at the University of Maryland,
Institute for Cyber-Enabled Research at Michigan State University,
and Astroparticle physics computational facility at Marquette University;
Belgium {\textendash} Funds for Scientific Research (FRS-FNRS and FWO),
FWO Odysseus and Big Science programmes,
and Belgian Federal Science Policy Office (Belspo);
Germany {\textendash} Bundesministerium f{\"u}r Bildung und Forschung (BMBF),
Deutsche Forschungsgemeinschaft (DFG),
Helmholtz Alliance for Astroparticle Physics (HAP),
Initiative and Networking Fund of the Helmholtz Association,
Deutsches Elektronen Synchrotron (DESY),
and High Performance Computing cluster of the RWTH Aachen;
Sweden {\textendash} Swedish Research Council,
Swedish Polar Research Secretariat,
Swedish National Infrastructure for Computing (SNIC),
and Knut and Alice Wallenberg Foundation;
European Union {\textendash} EGI Advanced Computing for research;
Australia {\textendash} Australian Research Council;
Canada {\textendash} Natural Sciences and Engineering Research Council of Canada,
Calcul Qu{\'e}bec, Compute Ontario, Canada Foundation for Innovation, WestGrid, and Compute Canada;
Denmark {\textendash} Villum Fonden, Carlsberg Foundation, and European Commission;
New Zealand {\textendash} Marsden Fund;
Japan {\textendash} Japan Society for Promotion of Science (JSPS)
and Institute for Global Prominent Research (IGPR) of Chiba University;
Korea {\textendash} National Research Foundation of Korea (NRF);
Switzerland {\textendash} Swiss National Science Foundation (SNSF);
United Kingdom {\textendash} Department of Physics, University of Oxford.